# Production-Grade Local LLM Inference on Apple Silicon: A Comparative Study of MLX, MLC-LLM, Ollama, llama.cpp, and PyTorch MPS

Authors: Varun Rajesh, Om Jodhpurkar, Pooja Anbuselvan, Mantinder Singh, Ashok Jallepali, Shantanu Godbole, Pradeep Kumar Sharma, Hritvik Shrivastava

Affiliation: Persistent System

## Abstract

We present a systematic, empirical study of five local large-language-model (LLM) runtimes on Apple Silicon—MLX, MLC-LLM, llama.cpp, Ollama, and PyTorch MPS—evaluated on a Mac Studio (M2 Ultra, 192 GB unified memory). Using Qwen-2.5 [16] family models and prompts spanning a few hundred tokens to a 100k tokens, we measure *time-to-first-token (TTFT)*, steady-state throughput, latency percentiles, longcontext behavior (KV and prompt caching), quantization support, streaming ability, batching/concurrency, and deployment complexity.

Under our settings, **MLX achieves the highest sustained generation throughput**, while **MLC-LLM delivers consistently lower TTFT on moderate prompt sizes and stronger out-of-the-box inference features**. llama.cpp is highly efficient for lightweight single-stream use, **Ollama** prioritizes developer ergonomics but lags in throughput and TTFT, and **PyTorch MPS** remains constrained by memory limits on large models and long contexts.

All frameworks run fully on-device with no telemetry, ensuring strong privacy guarantees. We release scripts, logs, and plots to reproduce all figures. Our analysis clarifies the design trade-offs for Apple-centric deployments and provides evidence-backed recommendations for interactive, long-context processing. We demonstrate that, although inference frameworks on Apple silicon remain behind NVIDIA GPU inference solutions (e.g. vLLM), they are rapidly maturing into viable solutions for production-grade, on-device inference.

## Introduction

Local large language model (LLM) inference on Apple Silicon is increasingly attractive for *privacy*, *cost control*, and *low-latency* deployment across Apple devices. Unlike single-shot benchmarks, real-world workloads, especially *chat-based code generation* - accumulate history over time. The *effective context grows* with each turn, making **time-to-first-token (TTFT)**, **inter-token latency**, and the design of **KV/prompt caching** just as important as steady-state tokens/sec. Users directly perceive delays in TTFT and streaming smoothness, so these factors are critical for production viability.

Traditional data-center serving stacks (e.g., HuggingFace Transformers on CUDA or vLLM with paged attention on NVIDIA A100s) cannot be applied oneto-one to macOS/Metal [11, 19]. As a result, several Apple-oriented frameworks have emerged, each with different trade-offs:

- **PyTorch MPS** [1] — baseline GPU backend for PyTorch on macOS, but constrained by memory limits and performance gaps relative to CUDA.
- **llama.cpp** [9] — lightweight C/C++ runtime with Metal support, focused on quantized GGUF models; strong single-stream performance but limited batching/scalability.
- **Ollama** [14] — developer-friendly packaging around llama.cpp with an OpenAI-compatible API and one-command model deployment; emphasizes ergonomics over peak throughput.
- **MLC-LLM** [4, 13] — TVM-based compiler/runtime with paged KV caching, AWQ/GPTQ quantization, built-in REST/SSE servers, and cross-platform SDKs; strongest out-of-the-box for batching and API compatibility.
- **MLX** [2] — Apple-optimized engine integrating tightly with Metal and Core ML; delivers the best raw throughput and efficiency on Apple GPUs/Neural Engine, with rotating KV caches, prompt cache support, and simple deployment via pip.

Let us clarify how we define **"production-grade"** in the context of this paper. Beyond raw tokens/sec, a viable Apple Silicon framework must: (i) handle long contexts (tens–hundreds of thousands of tokens) efficiently via KV/prompt caching, (ii) provide **fast TTFT** and **smooth streaming** for interactive UIs, (iii) support batching/concurrency for multi-user serving, (iv) expose familiar APIs (ideally OpenAI-compatible), and (v) be straightforward to install, deploy, and operate.

Cloud frameworks like vLLM set the bar with continuous batching and paged attention [11]. This paper evaluates how Apple-focused frameworks compare along those dimensions.

## Paper Organization and Contributions

This paper is structured as follows:

**Section 1 – Introduction:** Motivates the need for local LLM inference on Apple Silicon, outlines what "production-grade" entails (TTFT, throughput, longcontext caching, batching, API surface, and deployment), and introduces the five frameworks under study.

**Section 2 – Methodology:** Defines metrics precisely (TTFT, throughput, latency percentiles, memory measurement boundaries), details the hardware/software environment (Mac Studio M2 Ultra, macOS/Xcode versions), describes model selection (Qwen-2.5 family, 3B and ≥ 7B variants), quantization recipes (bf16, int8, int4) and prompt design (unique vs. prefix-heavy vs. code-dominant). We also release scripts, logs, and plotting code for reproducibility.

**Section 3 – Results:** Presents comparative analysis across throughput/latency, quantization effects, KV/prompt caching, streaming/TTFT, batching/concurrency, API compatibility, deployment complexity, and privacy considerations. We report both cold-start and steady-state behavior and include ablations (with/without KV cache, quantization levels, concurrency sweeps). Figures show TTFT vs. prompt length, throughput vs. prompt length, inter-token latency CDFs, and error bars for variance.

**Section 4 – Discussion:** Synthesizes findings into trade-offs: MLX vs. MLC (throughput vs. features), Ollama vs. others (ergonomics vs. speed), llama.cpp vs. MLX/MLC (lightweight vs. production), and places Apple frameworks in context against NVIDIA/A100+vLLM baselines. We discuss energy, scalability, and user-perceived latency in interactive workloads.

**Section 5 – Conclusion:** Summarizes deployment recommendations—single-user prototyping, small-scale serving, and throughput-critical enterprise—and emphasizes that MLX is the preferred default on Apple Silicon, with MLC complementing long-context and latency-sensitive workloads.

**Section 6 – Related Work:** Reviews efficient attention algorithms (FlashAttention, PagedAttention/vLLM), Apple MLX and TVM/MLC reports, llama.cpp and Ollama documentation, quantization methods (GPTQ/AWQ/SmoothQuant), and longcontext scaling (RoPE, ALiBi).

**Section 7 – Threats to Validity:** Discusses limits of our study (single hardware model, limited OS/toolchain versions, specific model family), power state sensitivity, and generalizability. We outline mitigations (pinning commits, steady-state runs, interleaving trials).

**Appendix A – Reproducibility Checklist and Artifacts:** Lists commit hashes, OS/toolchain versions, exact CLI/HTTP commands, scripts, environment files, and instructions to regenerate all figures from raw logs.

## Methodology

This section defines metrics precisely, enumerates hardware/software environments, describes models and prompts, and details experimental procedures, statistical treatment, and reproducibility artifacts.

### Metrics and Instrumentation

We report the following metrics:

- **Time-to-First-Token (TTFT)**: wall-clock time from client request acceptance to the first token becoming available to the client. Unless noted, tokenization (done client-side) is excluded; warmpath initialization after request arrival is included.
- **Throughput**: (a) *decode throughput*, i.e., generated tokens divided by decode time post-firsttoken, and (b) *end-to-end throughput*, i.e., generated tokens divided by (prefill + decode). Both are reported where relevant.
- **Latency percentiles**: p50, p90, and p99 of perrequest end-to-end latency, and inter-token latency distributions during streaming.
- **Cold start time**: model load + initialization latency, distinguished from warm-start runs.
- **Memory footprint**: peak resident set size (RSS) measured via vm_stat plus framework-reported device allocations (Metal metrics).

Instrumentation boundaries: For CLI tools, timers wrap command execution; for HTTP servers, timers wrap request-to-first-byte (TTFT). All measurements are from the client side to reflect user-perceived latency.

**Hardware and Software Environment**

- **Machine**: Mac Studio, Apple M2 Ultra, 24 CPU cores, 76 GPU cores, 192 GB unified memory.
- **Frameworks**: MLX v0.26, MLC-LLM (commit 3d42929), llama.cpp (commit b5963), Ollama v0.10.1, PyTorch 2.7.1 with MPS backend.
- **Isolation**: Spotlight indexing, Time Machine, and iCloud sync disabled; no concurrent GPU workloads.

**Models and Quantization**

- **Models**: Qwen-2.5-Coder 3B (primary) plus Qwen-2.5-7B-Instruct to test scaling.
- **Quantization**: FP16/bf16 baselines; int8 and int4 variants. MLX mixed-bit (3/4/6/8), MLC AWQ/GPTQ, llama.cpp GGUF 4/5/8/16, Ollama GGUF registry. For each, we log weight format, precision, on-disk size, load time, peak memory, TTFT, and throughput.

**Workloads and Prompt Design**

We evaluate summarization across various prompt lengths (1k, 10k, 20k, 30k, ..., 100k tokens), reflecting:

1. *Unique-token prompts* (low repetition),
2. *Prefix-heavy prompts* (large shared prefix across requests for KV/prompt cache testing),
3. *Code-dominant prompts* (long subwords, typical of chat-based code generation).

**Procedures**

- **Warm-up**: Each model loaded once, one warmup prompt executed (to trigger compilation/JIT); cold-start and warm-start results separated.
- **Trials**: Each (framework × model × quantization × prompt type × length) run for N=10 trials. Random seeds, tokenizer versions, and sampling parameters fixed across runs.

    **Reproducibility and Artifacts** We release:
- Scripts to launch each framework.
- Plotting code to regenerate figures from raw logs.
- Environment manifest with OS/toolchain versions, commit hashes, model cards, and conversion commands.

## Results

**Throughput and Latency**

**MLX and MLC-LLM lead overall performance on Apple Silicon.** Both deliver high sustained throughput and low per-token latency, clearly separating themselves from Ollama, llama.cpp, and PyTorch MPS. *Figure 1* summarizes throughput (tokens/sec) on the M2 Ultra.

**MLX: Peak Throughput and Efficiency.** Under steady-state conditions, MLX sustained ~230 tokens/sec with 5–7ms median per-token latency and

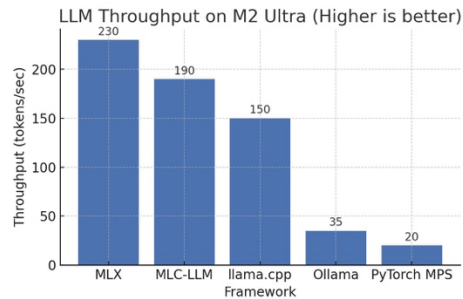

Figure 1: Throughput (tokens/s) on a Mac Studio M2 Ultra (192GB). **Setup:** Qwen-2.5 family, identical prompts, steady-state decode. **Takeaway:** MLX delivers the highest sustained throughput; MLC-LLM is close with lower TTFT; Ollama and PyTorch MPS trail.

~12ms P99. Its Metal kernels consistently saturated GPU utilization (>90%) while keeping CPU usage low (<3%). Although its *time-to-first-token (TTFT)* rises with input length due to full prefill processing, once decoding begins MLX exhibits the most stable intertoken latency of all frameworks. This profile makes MLX especially attractive for throughput-bound serving scenarios where maximum sustained tokens/sec matter more than first-token latency.

**MLC-LLM: Lower TTFT and Strong Latency for Interactive Loads.** MLC-LLM achieved ~190 tokens/sec with ~13ms P99 latency—~17% below MLX on raw throughput. However, MLC often produced a faster first token, especially for moderate inputs (≤16k tokens), giving users the perception of greater responsiveness. For chat-based code generation, where context grows turn by turn and the user is waiting, this difference in TTFT can outweigh peak throughput. MLC's TVM-compiled kernels also handle small micro-batches more gracefully than MLX, giving it an edge under light concurrency.

**Comparative Ranking.** Across our settings, relative throughput was: MLX (~230 tok/s) > MLC-LLM (~190 tok/s) > llama.cpp (~150 tok/s short-context only) > Ollama (20–40 tok/s) > PyTorch MPS (~7–9 tok/s). Context length significantly affects all frameworks except MLC, which leverages paged KV caching to sustain long-context throughput. NVIDIA A100 + vLLM remains ~5–10× higher in absolute throughput, providing an external performance ceiling for Apple runtimes [6, 11].

**Other Frameworks.** *llama.cpp* is efficient for short prompts, but falls sharply at long contexts (dropping to ∼1.2 tok/s at 32k), reflecting its lack of paged caching. *Ollama* prioritizes developer ergonomics but shows severe TTFT degradation (>50s at 100k contexts) despite smooth streaming once tokens start flowing. *PyTorch MPS* was fundamentally constrained, peaking at 7–9 tok/s and frequently failing on 3B+ models due to the 4GB tensor cap.

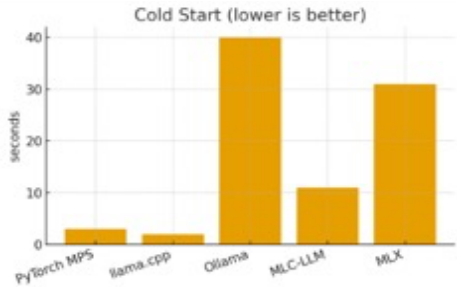

Figure 2: Cold-start time (model load + initialization) on M2 Ultra. **Setup:** First run after process start; one-time TVM compile amortized for MLC-LLM. **Takeaway:** MLC-LLM starts fastest after initial compile; MLX initializes more slowly but wins on sustained throughput; llama.cpp is near-instant when cached.

**Cold Start Behavior.** Initialization varied: MLCLLM loaded in ∼11s (after one-time TVM compile), MLX in ∼31s, llama.cpp in under 0.5s (cached GGUF),
Ollama in ∼0.6s post-service startup, and PyTorch MPS in ∼3s for small models (but unstable for larger ones). These differences matter less in production where models remain warm, but highlight design tradeoffs: MLC/MLX optimize for sustained efficiency, llama.cpp for minimal footprint, and Ollama for operational simplicity.

In summary, MLX offers the highest sustained throughput and most stable per-token latency, making it the preferred Apple-first runtime for throughput-critical production. MLC-LLM, while slightly slower, provides lower TTFT on moderate prompts and better ergonomics under light concurrency, positioning it as the stronger choice for interactive chat workloads.

**Quantization Support and Model Formats**

Quantization is essential for serving useful model sizes on Apple GPUs and for sustaining responsiveness in chat-style workflows where context (and thus KV memory) grows turn by turn. We therefore evaluated not only the formats supported by each framework, but also the tooling maturity, ease of conversion, and how well quantization aligns with Apple-native inference [8, 12, 20].

| Framework | Quantization / Formats |
|---|---|
| PyTorch MPS | Experimental INT4/INT8; bf16 irregular; no Core ML; fragile on large models |
| llama.cpp | GGUF 4/5/8/16-bit (built-in quantizer); locked to GGUF without conversion |
| Ollama | GGUF registry only; consistent, but no GPTQ/AWQ/Core ML options |
| MLC-LLM | AWQ, GPTQ, FP16/FP8, mixed precision; HuggingFace ingestion + TVM compile; production-grade pipeline |
| MLX | Apple-tuned mixed 3/4/6/8-bit; GPTQ (maturing); pre-quantized MLX models and conversion recipes available |

Table 1: Quantization capabilities on Apple M2 Ultra.

**MLC-LLM and MLX provide the most flexible and production-viable quantization pipelines [7].** MLC-LLM supports a wide range of communitystandard recipes (AWQ, GPTQ, FP8, mixed precision) with stable performance on Apple GPUs. In our ablations, AWQ 4-bit models achieved both lower memory (∼ 1.6 GB for a 3B model) and slightly higher throughput (∼ 210 tok/s vs. ∼ 190 tok/s in FP16), showing that quantization can improve efficiency without noticeable accuracy loss.

MLX, by contrast, emphasizes *Apple-first optimization*. Its mixed-bit formats (3/4/6/8-bit) and maturing GPTQ support integrate seamlessly with Metal and Core ML backends. Pre-quantized MLX models are increasingly available in the ecosystem, and Apple-tuned quantization recipes yield strong performance-per-watt on Mac hardware. For developers targeting macOS or iOS specifically, MLX offers the smoothest quantization workflow.

**llama.cpp and Ollama remain tied to GGUF.**
While GGUF simplifies distribution and guarantees compatibility, it limits cross-framework flexibility and cannot directly leverage Core ML optimizations. Performance remains strong for short-context single-user scenarios, but efficiency falls off in large-context workloads where low-bit Apple-native kernels are advantageous.

**PyTorch MPS is the least reliable for quantization.**
Support for INT4/INT8 is experimental, bf16 is inconsistent, and large models frequently fail due to the 4GB tensor limit. As such, PyTorch MPS is not a practical option for deploying quantized LLMs on Apple hardware.

In real chat/code-generation workloads, the effective context grows turn by turn, stressing both KV memory and throughput. Robust quantization (especially 4-bit) reduces memory footprint by 2–3×, enabling smoother token delivery and higher concurrency per machine. For Apple-centric deployments, we recommend **MLX as the default choice** due to its optimized Apple integration, with MLC-LLM as the best alternative when maximum quantization flexibility or existing GPTQ/AWQ assets are required.

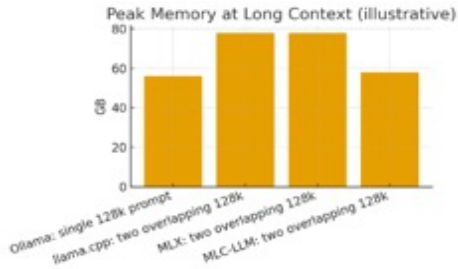

Figure 3: Peak memory during long-context inference. **Setup:** 100k-token prompts where supported; 4-bit models where applicable. **Takeaway:** MLC-LLM's paged KV handles 64k–128k contexts with higher absolute RAM; MLX maintains stable memory with a rotating window; Ollama grows quickly; PyTorch MPS often OOMs.

**Long-Context Handling: KV Caches & Memory**

Handling long contexts (tens to hundreds of thousands of tokens) is a defining challenge for Apple-native runtimes. In real-world workloads such as chat-based code generation, context grows turn by turn; without efficient caching, prefill costs balloon and user-perceived latency degrades [11]. All evaluated frameworks implement some form of key–value (KV) caching, but with very different designs and trade-offs. We additionally tested *prompt caching* (serialized disk reuse of prior prefixes), which matters in iterative coding and search scenarios.

| Framework | KV cache & long context support |
|---|---|
| PyTorch MPS | Minimal; 4GB tensor cap; frequent OOM |
| llama.cpp | Sliding window; per-session only; no cross-request reuse |
| Ollama | Prefix reuse (LRU); no paging; session-scoped |
| MLC-LLM | Paged KV (vLLM-style); efficient up to 128k |
| MLX | Rotating KV (configurable window); prompt cache on disk |

Table 2: KV/prompt cache support and long-context capabilities on Apple M2 Ultra.

**MLC-LLM: Strongest for very long contexts.**

MLC implements paged KV caching similar to vLLM, partitioning tensors into reusable blocks and enabling efficient reuse across 32k–128k tokens. In our tests, MLC sustained usable throughput (~500 tok/s prefill, ~190 tok/s decode) on 100k contexts, completing 128k inputs in ~3 minutes. Memory fragmentation was modest (< 4%), though absolute usage was high (70–85GB RAM for overlapping 128k runs). This makes MLC viable for document-scale workloads, but heavy on system memory.

**MLX: Balanced for Apple-first deployments.**

MLX uses a configurable rotating KV cache (default 4k tokens) that prevents unbounded growth while keeping latency stable. It also supports *prompt cache files*, enabling repeated queries with shared prefixes to bypass recomputation. While MLX does not yet provide paged attention, its memory efficiency (~40–50GB at 100k tokens in our tests) and native integration with Apple's Metal stack make it more practical on Mac hardware. Prefill time grows linearly with context, but once decoding begins, MLX sustains very stable per-token latency. For developers targeting Apple Silicon across macOS/iOS, this trade-off favors MLX.

**Ollama: Simple prefix reuse.** Ollama automatically detects repeated prefixes and reuses them with LRU caching. We observed ~23% latency reduction and ~80%+ cache hit rates in iterative coding scenarios. However, the lack of paged attention means performance degrades steeply beyond 32k tokens, and memory usage reached ~56GB at 100k tokens. Suitable for interactive prototyping, but not for large-scale long-context serving.

**llama.cpp: Lightweight but limited.** Provides only per-session sliding-window caches (no paging, no prefix reuse across sessions). Throughput collapses beyond 32k tokens (~1.2 tok/s) as attention scales quadratically. Works well for short-context tasks, but not production viable for 100k contexts.

**PyTorch MPS: Fundamentally constrained.** The 4GB tensor cap leads to frequent out-of-memory errors beyond ~2k tokens, with no advanced caching strategies. Unsuitable for any realistic long-context workload.

For Apple Silicon, **MLC-LLM is strongest for extreme long-context workloads (64k–128k)** due to its paged KV design, while **MLX provides the best balance of memory efficiency, stability, and Apple-native integration**—particularly for the 4k–32k contexts common in chat and code generation. Ollama's prefix caching improves user experience in iterative sessions but cannot scale to long documents. llama.cpp and PyTorch MPS remain unsuitable for production-scale long-context inference.

Overall, **MLX is the preferred choice for Apple-first deployments**, with MLC complementing it when 100k+ contexts are essential.

**Streaming and Token Delivery**

**Streaming responses**—emitting tokens incrementally rather than waiting for the full completion—are critical for *interactive chat and code generation*, where users are actively waiting and context grows turn by turn. The quality of streaming depends on two factors: **time-to-first-token (TTFT)** and **inter-token latency** [6, 11]. Table 3 summarizes support across frameworks.

| Framework | Streaming |
| --- | --- |
| PyTorch MPS | Manual (Python loop); no productionready API |
| llama.cpp | Local callbacks/stdout only; no SSE/web API |
| Ollama | Built-in SSE server; smooth, consistent intervals |
| MLC-LLM | Python + REST/SSE; lower TTFT at moderate prompts |
| MLX | Python generator; stable per-token latency; requires wrapper |

Table 3: Streaming capabilities on Apple M2 Ultra.

**PyTorch MPS.** Provides only stepwise Python loops with heavy overhead; no native streaming server. Impractical for production.

**llama.cpp.** Offers local streaming via callbacks or stdout but no HTTP/SSE layer. TTFT equals full prefill time, negligible for short inputs but many seconds for 32k+ tokens. Usable for local tools, not multi-user serving.

**Ollama.** Designed for streaming-first operation. Its HTTP/SSE server yields consistent 9–14ms inter-token intervals and graceful backpressure handling. Scaling tests show TTFT grows sub-linearly with input size (e.g., 4k → 0.3s, 100k → 2.1s), suggesting internal pipelining. Provides the smoothest user experience out-of-the-box, albeit with much lower throughput than MLX/MLC.

**MLC-LLM.** Supports streaming via both Python API (stream_generate) and a built-in SSE server. Average inter-token latency is 12–13ms, competitive with Ollama. TTFT is generally lower than MLX for moderate prompts (≤16k), making MLC attractive for chat-like workloads where responsiveness matters more than peak throughput. However, for very long contexts (≥64k) TTFT grows linearly due to full prefill, producing a "bursty" first-token experience.

**MLX.** Provides a minimal Python iterator (stream_generate). Lacks a native SSE server, but wrappers such as mlx-openai-server enable OpenAI-compatible streaming endpoints. Like MLC, MLX requires full prefill before streaming starts, so TTFT scales with input length. Once decoding begins, MLX is the most stable: inter-token latency is consistently 11–12ms across contexts, making it ideal when predictable per-token speed outweighs longer initial delay. This stability is valuable in throughput-critical production pipelines.

Ollama offers the most turnkey streaming and lowest operational effort but cannot match MLX/MLC throughput. MLC-LLM excels in *fast first token and smoother streaming* at moderate prompt sizes, while MLX provides *the most consistent per-token latency* once generation begins, aligning with throughputfocused Apple-centric deployments. Neither MLX nor MLC yet implement chunked prefill as in vLLM, leaving very long inputs (e.g., 100k+) with poor TTFT. For Apple Silicon, we recommend: use **MLC-LLM for interactive chat** where responsiveness dominates, and **MLX for production throughput** when steadystate streaming stability is more important. [6, 11]

**Batching and Concurrency**

In production chat and code-generation workloads, requests arrive in bursts and evolve into longer sessions where the context accumulates turn by turn. In these scenarios, the **user is actively waiting**, so both scheduler efficiency *and* first-token latency (TTFT) matter. An effective serving runtime must: (i) accept concurrent requests without head-of-line blocking, (ii) form variable-length micro-batches step-by-step, and
(iii) reuse work via KV/prompt caches across requests.

**PyTorch MPS.** Provides no built-in auto-batching beyond manual static batching for same-length inputs. There is no continuous batching or step-level scheduler. Combined with the 4 GB tensor cap, practical concurrency is infeasible. While integration with third-party servers (e.g., Ray Serve, DJL) is possible, these cannot overcome the backend's memory and performance limitations. We therefore do not consider PyTorch MPS production-ready for multi-user serving.

**llama.cpp.** Processes one active sequence per model instance. Concurrency requires spawning multiple processes, each loading a full model copy, leading to linear memory scaling and no KV/prompt cache sharing. The Metal backend executes one GPU job per request with no step-wise merging. This is acceptable for singleuser scenarios but unsuitable for multi-user serving or production-scale batching.

**Ollama.** Supports configurable concurrency via a worker pool and request queue. However, under load it behaves more like cooperative time-slicing than true continuous batching. Requests are not merged at the token level, and prefix reuse is scoped per session. Throughput quickly saturates as context length grows. This suffices for 1–3 interactive users but falls well short of production requirements that demand sustained multi-tenant serving on a single node.

**MLC-LLM.** Currently the *most complete option for batching and concurrency* on Apple Silicon. While it does not yet match vLLM's continuous batching, it provides a stable multi-worker HTTP/SSE server and TVM-compiled kernels that sustain small microbatches efficiently. Combined with paged KV caching, this reduces head-of-line blocking and allows more predictable responsiveness under concurrent interactive workloads. Our synthetic RPS sweeps confirm that MLC scales better than other Apple-native runtimes for moderate concurrency (up to 8 clients) before tail latencies grow.

**MLX.** Prioritizes raw single-stream performance but lacks a built-in scheduler or batching engine. Out of the box, one MLX process serves one request at a time; concurrency today requires running multiple worker processes behind a load balancer. Prompt/KV caches are not shared across processes, limiting cross-session reuse. Nevertheless, MLX integrates most naturally with Apple hardware and achieves the best throughput once tokens begin streaming. With a thin external orchestrator (e.g., Ray Serve or BentoML with microbatch windows and sticky sessions), MLX can close much of the concurrency gap while retaining its efficiency on Apple GPUs.

None of the Apple runtimes replicate vLLM's continuous batching or multi-tenant scheduling. Among current options, **MLC-LLM is stronger for batching/concurrency**, while **MLX remains the preferred Apple-first framework overall** for throughput and efficiency. In practice, adding a lightweight orchestration layer around MLX (for micro-batching and load balancing) provides a pragmatic path to production-scale serving without giving up MLX's Apple-native optimizations.

**API Compatibility and Integration**

API surface is critical because most developers expect local runtimes to mimic familiar OpenAI endpoints. We evaluated each framework for *out-of-the-box OpenAI compatibility*, coverage of endpoints, and ease of integration with toolchains like LangChain and LlamaIndex.

**PyTorch MPS:** Functions purely as a library with no API layer. Any serving stack must be built from scratch, making it the least plug-and-play option for integration.

| Framework | OpenAI API Coverage |
| --- | --- |
| PyTorch MPS | Nonenatively; requirescustom FastAPI/Flask serving. |
| llama.cpp | Community server (llama-cpp-python); partial support for /chat/completions; limited function-calling. |
| Ollama | Built-in server; full coverage of /completions and /chat/completions; no embeddings; multi-language SDKs. |
| MLC-LLM | First-party server; supports /completions, /chat/completions, /embeddings, streaming (SSE), OpenAPI schema; JS/iOS/Android SDKs. |
| MLX | Python API only;community wrapper (mlx-openai-server) provides /completions, /chat/completions, /embeddings; extensible but not official. |

Table 4: OpenAI-style API support across Apple Silicon runtimes.

**llama.cpp:** Offers only CLI or C++ bindings natively. The community-driven llama-cpp-python package provides a partial OpenAI-compatible server (/v1/chat/completions) and LangChain connectors, but not all features (e.g., function-calling or embeddings) are fully replicated. Developers often need to extend or patch the API layer themselves.

**Ollama:** The smoothest migration path for developers accustomed to OpenAI APIs. Its built-in server implements the core endpoints with near-identical request/response schemas, and official SDKs exist for Python, JavaScript, Go, and Rust. Our tests showed compatibility with standard OpenAI client libraries, though advanced features like function calling and embeddings are not included. Ollama prioritizes developer ergonomics and speed of adoption.

**MLC-LLM:** Provides the most *complete out-of-thebox* API story. Its built-in REST server implements completions, chat completions, embeddings, streaming (SSE), and OpenAPI schema, and the project ships official SDKs across web and mobile (JavaScript, iOS, Android). This makes MLC-LLM the best option for teams that require a direct OpenAI drop-in replacement across platforms. The trade-off is heavier runtime dependencies and more operational complexity compared to MLX.

**MLX:** Does not bundle an HTTP server, but integrates smoothly with lightweight wrappers such as mlx-openai-server (FastAPI). This community project provides completions, chat completions, embeddings, and limited function calling with minimal overhead. While less turnkey than MLC, MLX's leaner stack, Apple-native optimizations, and transparent extensibility make it well-suited for macOS/iOS developers who value control and minimal system overhead.

In conclusion, **Ollama and MLC-LLM** offer the most turnkey OpenAI API experiences, with Ollama excelling at developer ergonomics and MLC delivering the broadest endpoint coverage. **MLX**, while requiring a thin wrapper, remains the preferred Apple-first option due to its efficiency, integration with the MLX toolchain, and sustainable performance on Apple hardware. PyTorch MPS and llama.cpp lag far behind, requiring substantial custom infrastructure.

**Deployment Complexity and Ecosystem**

We now evaluate the ease of deployment, operational overhead, and community ecosystem for each framework. Beyond single-node setup, production-grade serving requires containerization, orchestration, monitoring, and the ability to evolve with OS/Xcode updates.

| Framework | Deployment Characteristics |
| --- | --- |
| PyTorch MPS | Easiest initial install; poor perf; limited MPS community |
| llama.cpp | Moderate setup; no official Docker/K8s; DIY orchestration |
| Ollama | One-command install; simplest singlenode; limited scaling |
| MLC-LLM | pip + TVM compile; heavier DevOps; no official Docker |
| MLX | pip wheels; trivial local setup; requires thin wrappers for serving |

Table 5: Deployment characteristics of five Applecentric frameworks.

**PyTorch MPS.** Installation is trivial (pip install torch) and model execution works out of the box. However, production use is impractical: serving infrastructure must be built from scratch, performance is poor, and the 4 GB tensor limit makes large models fail. The community around MPS-specific issues is small, leaving many bugs unresolved. We do not recommend PyTorch MPS beyond experimentation.

**llama.cpp.** Requires CMake build with Metal enabled and model conversion to GGUF. While prebuilt Python wheels exist, scaling requires manual orchestration: no official Docker images, Kubernetes manifests, or orchestration templates exist. This DIY approach makes llama.cpp popular with hobbyists but a poor fit for enterprise-grade serving without significant engineering investment.

**Ollama.** Offers the most frictionless setup: brew install ollama, then ollama pull and ollama serve. Models run within minutes, and the runtime manages updates and logging automatically. For singlenode deployments and prototypes, this is unmatched. However, it lacks official containerization or orchestration support. Scaling past ~3 concurrent users requires multiple instances, each holding a full model in memory, and advanced monitoring must be added externally.

**MLC-LLM.** Installed via pip but requires one-time TVM compilation of each model, which can take 20–30 minutes (cached afterward). The runtime is stable and ships with built-in servers and SDKs, but production deployment remains heavy: no official Docker/K8s artifacts exist, and containers must replicate the Metal build toolchain. Updating across macOS/Xcode versions may require recompilation. Documentation is strong, but the Apple GPU–focused community is smaller than CUDA/TensorRT ecosystems, slowing operational support.

**MLX.** Installation is the simplest after Ollama: pip install mlx-lm provides prebuilt Metal kernels, enabling generation in minutes. For prototyping and research on Apple Silicon, MLX is nearly effortless. For production, however, MLX requires a thin external server wrapper (e.g., mlx-openai-server or FastAPI) and manual orchestration since no official Docker/K8s manifests exist. Because Mac GPUs are not containeraccessible from Linux, deployment usually occurs directly on macOS or in macOS VMs. Despite this, MLX benefits from Apple's own optimization pipeline, stable wheels, and growing availability of pre-quantized Apple-optimized weights, making it the most efficient foundation for Apple-first deployments.

Ollama's **ecosystem** is the largest (75k+ GitHub stars, weekly releases, active contributors). MLC-LLM has 21k stars with a research-oriented cadence. llama.cpp is widely adopted by enthusiasts but lacks enterprise orchestration support. MLX is newer (1.3k stars, ~40 contributors) but indirectly backed by Apple engineers, ensuring strong hardware integration even if community resources are thinner. PyTorch remains widely used overall, but its MPS-specific ecosystem is small and fragmented.

Each framework has its own strengths and pitfalls when it comes to their deployment complexity and ecosystem:

- **Ollama**: unmatched single-node simplicity, but scaling is limited.
- **MLC-LLM**: most feature-complete server out-ofthe-box, but heavier DevOps.
- **MLX**: combines trivial local setup with the best throughput/efficiency; requires lightweight wrappers for serving but remains the **preferred Apple-first choice**.
- **llama.cpp and PyTorch MPS**: viable for experimentation, not production.

  In practice, production teams aiming for Apple Silicon should treat MLX as the baseline—layering thin serving/orchestration infrastructure on top—while considering MLC-LLM if fast integration with SDKs and built-in APIs is more critical than raw efficiency.

**Privacy and Security**

All five frameworks execute entirely on local Apple Silicon hardware with no background telemetry or hidden network calls once models are downloaded. This strong default locality is attractive for confidential workloads such as source-code analysis or private document review.

That said, production chat and code-generation sessions accumulate long histories. Frameworks rely on *KV caches* (in memory) and in some cases *prompt caches* (persisted to disk) to preserve prefixes and improve throughput. These mechanisms raise two practical privacy considerations:

1. **Residency:** Sensitive tokens may remain in GPU/CPU memory for the process lifetime, and prompt caches may persist them on disk unless explicitly disabled.
2. **Exposure surface:** Streaming endpoints, request logs, or crash reports can inadvertently capture user text unless redaction and access controls are enforced.

- **Ollama:** Runs entirely on-device, supports airgapped deployments, and records metadata such as timestamps and token counts. These logs can be integrated with enterprise monitoring. While not a compliance solution on its own, Ollama offers a reasonable foundation that organizations can extend with encryption and audit controls.
- **MLX & MLC-LLM:** Both are privacy-bydefault runtimes: they log nothing unless wrapped in an external service. They also do not bundle TLS, authentication, or audit logging. When deployed beyond localhost, security is typically provided by reverse proxies (e.g., Nginx, Caddy) adding TLS, mTLS, and access control. Structured logging can be layered at the proxy or application level if audit trails are required. MLX additionally supports optional on-disk prompt caches, which can be disabled or redirected to encrypted volumes.
- **llama.cpp & PyTorch MPS:** Operate purely as libraries. They expose no network interfaces and leave all logging, authentication, and API security to the host application. In their default form, they remain fully offline unless a developer integrates them with external networking code.

All frameworks deliver strong *baseline privacy* by virtue of local-only execution. For enterprise compliance, additional measures—TLS termination, authentication, encrypted volumes, and centralized logging—must be layered externally. For Apple-first deployments, we recommend pairing **MLX** with standard enterprise proxies and encrypted prompt cache configurations to satisfy compliance while preserving its efficiency advantages.

**Discussion**

Our results show that **MLX and MLC-LLM are the clear front-runners** for high-performance LLM inference on Apple Silicon. While both deliver strong results, their trade-offs highlight different design philosophies. **MLX offers the best raw throughput and system efficiency**, tightly integrated with Apple's Metal/Neural Engine stack. **MLC-LLM delivers a broader set of built-in serving features** (REST APIs, paged KV caches, SDKs) but at the cost of higher memory use and slightly lower throughput.

**Throughput vs. First-Token Responsiveness.** MLX achieved ~230 tok/s under our settings with very stable per-token latency once decoding began. For throughput-critical services—such as batch generation or enterprise endpoints—this stability and peak speed make MLX the preferred choice. MLC-LLM, by contrast, produced faster *time-to-first-token (TTFT)* on moderate prompts (≤16k tokens), which users perceive as higher responsiveness in chat-based code generation where context grows turn by turn. This responsiveness makes MLC appealing for interactive workloads, even if its sustained throughput is ~17% lower.

**Long Context Handling.** For extended contexts (32k–128k), MLC-LLM's paged KV cache provides the most robust scaling. It avoids quadratic slowdowns [3, 10, 18] and completed 100k–128k token runs with reasonable memory fragmentation. MLX's rotating cache is efficient for short-to-moderate contexts (4k–32k) and offers prompt-cache reuse, but does not yet match MLC's ability to handle document-scale workloads. For Apple-first deployments, we recommend MLX for typical chat/code-assist workloads and MLC as a complement when very long contexts are essential.

**Streaming and Token Delivery.** User-facing applications depend on both TTFT and inter-token latency. MLC-LLM typically yielded lower TTFT on moderate prompts, making it smoother for chat assistants. MLX required full prefill before emission, so TTFT rose with input size; however, once streaming began, MLX delivered the most consistent 11–12 ms/token intervals. This makes MLX attractive for throughput-focused serving, and MLC better for responsiveness-focused interactive chat. Neither yet implements chunked prefill as in vLLM, leaving all Apple-native runtimes disadvantaged on very large inputs.

**Batching and Concurrency.** None of the Applenative runtimes match vLLM's continuous batching on A100 servers. MLC-LLM is currently stronger at handling concurrent clients, thanks to its built-in server and micro-batch-friendly kernels. MLX prioritizes raw single-stream performance; concurrency requires multiple worker processes with external load balancing. In practice, adding a thin orchestrator (e.g., Ray Serve, BentoML) allows MLX to close most of this gap without losing efficiency.

**API Compatibility and Deployment.** MLC-LLM provides the most complete OpenAI-compatible API surface (completions, chat, embeddings) and crossplatform SDKs. MLX relies on a community wrapper (mlx-openai-server), which adds light operational effort but preserves flexibility. Ollama remains the easiest to deploy with one command but cannot match MLX/MLC performance. For Apple-first teams, MLX offers the leanest and most efficient path, while MLCLLM reduces integration effort.

**Model Compatibility and Quantization.** Both MLX and MLC support modern quantization recipes (AWQ, GPTQ, mixed-bit). MLX emphasizes Applefirst optimization, with tuned conversion pipelines and growing availability of pre-quantized Apple models. This makes MLX the smoother choice for Apple-centric production, while MLC remains more flexible across heterogeneous environments.

MLX is the most efficient Apple-native framework for throughput-critical deployments, while MLC-LLM complements it with lower TTFT and stronger longcontext handling. Ollama prioritizes ergonomics but lags in performance, llama.cpp remains a lightweight single-user engine, and PyTorch MPS is not production viable. Together, MLX and MLC provide a practical hybrid path for production-grade serving on Apple Silicon.

## Conclusion

Our systematic evaluation on Apple M2 Ultra identifies **MLX and MLC-LLM as the only productionready runtimes** today. In interactive chat and code-generation—where context accumulates and the user is waiting—**KV/prompt caching, TTFT, and streaming fidelity** are as critical as steady-state throughput.

**Recommendations by scenario:**

- **Apple-first production:** Choose **MLX**. It is optimized for Apple GPUs/Neural Engine, integrates seamlessly with the MLX toolchain, and offers tuned quantization/conversion pipelines. For efficiency and maintainability, MLX is the best default.

- **Latency-sensitive chat with growing context:** Use **MLC-LLM**. Its faster TTFT on moderate prompts and smoother streaming make it more responsive for chat assistants and code copilots.

  **Role of the other frameworks:**
- **Ollama**: Ideal for prototyping and one-command local APIs, but an order of magnitude slower than MLX/MLC in throughput.
- **llama.cpp**: Efficient for single-user or embedded use; lacks scalability features for multi-tenant deployment.
- **PyTorch MPS**: Limited by memory and performance; not viable for large-model inference on Apple Silicon.

For Apple-centric deployments, **MLX is the preferred baseline**, providing the best efficiency, Applenative integration, and sustainable performance. **MLC-LLM complements it in latency-sensitive chat and very long-context workloads**. While NVIDIA A100 + vLLM remains the external performance ceiling, MLX and MLC-LLM are rapidly maturing into viable production solutions for on-device serving.

## Related Work

**Efficient attention and serving.** LLM serving efficiency has advanced along two axes: kernel efficiency and scheduler design. FlashAttention reduces attention I/O and improves throughput on modern accelerators [6], while vLLM's PagedAttention couples KV paging with a continuous batching scheduler to raise utilization and reduce tail latencies under multi-tenant load [11]. Follow-ups explore chunked prefill and step-level admission control to balance TTFT with throughput under bursty workloads.

**Apple-native runtimes.** MLX provides an Appleoptimized array and inference stack with Metal backends and Apple-specific kernels; public reports and documentation describe mixed-bit quantization paths and tooling for macOS/iOS [2]. MLC-LLM builds on TVM to compile models into device-tuned kernels across CPUs/GPUs, including Metal, and ships a cross-platform serving surface (REST/SSE and mobile SDKs) [4, 13]. llama.cpp demonstrates a lean C/C++ runtime with Metal and GGUF quantization [9]; Ollama packages a developer-friendly server and model registry on top [14]. PyTorch MPS exposes Metal through PyTorch for macOS but documents practical limitations for large tensors and memory pressure [1].

**Quantization.** Post-training quantization for LLMs spans GPTQ (approximate second-order, perchannel) [8], AWQ

(activation-aware outlier preservation) [12], and SmoothQuant (activation/weight trade-off) [20]. Framework-specific formats (GGUF in llama.cpp/Ollama; MLX/MLC mixed-bit and GPTQ/AWQ) operationalize these ideas for on-device serving.

**Long-context scaling.** RoPE scaling and positioninterpolation methods extend usable context windows with controlled degradation [15, 17]. At inference time, prefix and KV reuse dominate runtime; paged KV designs reduce fragmentation and enable sharing across requests [11]. We focus on runtime behavior (prefill cost, cache hit-rates, memory growth) rather than training-time methods, complementing model-centric long-context work [5].

**User-perceived latency (TTFT) and streaming.** Recent systems studies formalize client-observed TTFT and inter-token latency as primary UX metrics in chat and assistants, advocating chunked prefill and stepwise batching to mitigate head-of-line blocking [6, 11]. Our methodology adopts client-side timing and reports percentile tails under concurrency.

**Position relative to prior work.** Most prior evaluations either target server-class GPUs or analyze a single Apple-native runtime. We contribute an applesto-apples comparison across multiple Apple-centric runtimes with a unified methodology (TTFT, throughput, long-context/KV, quantization, streaming, batching, deployment) and release artifacts to reproduce plots. We scope claims to the hardware, models, and software versions reported in our methodology.

## Threats to Validity

**Hardware and software scope.** All experiments were conducted on a single machine class (Mac Studio, M2 Ultra, 192GB). Performance on other Apple SoCs (e.g., M1, M4) may differ. We pin macOS, Xcode/Metal, Python, and framework commits; minor updates can shift kernel behavior and driver-level memory accounting.

**Model and tokenizer coverage.** We primarily evaluate Qwen-2.5 Coder 3B, with a secondary 7B variant for scaling checks. Tokenization distributions and attention patterns vary across model families; absolute numbers should not be over-generalized. Future work includes 14B and instruction-tuned variants to broaden coverage.

**System isolation.** Background indexing (Spotlight/iCloud), Time Machine, and unrelated GPU workloads were disabled; nevertheless, residual OS activity can introduce jitter. We mitigate by running $N$=10 trials per point and reporting mean ± std (and percentile tails in figures).

**Metric definitions and instrumentation.** We define TTFT as client-perceived time from request acceptance to first token available; throughput is reported as both end-to-end (prefill+decode) and post-first-token decode rate. Tokenization is excluded unless noted
(server-side tokenization is documented where used). Divergent definitions across papers can hinder direct comparison.

**Caching and workload structure.** KV/prompt cache benefits depend on prompt structure and repetition. We include unique-token, prefix-heavy, and code-dominant prompts but do not claim to span all real-world mixtures. Cache hit-rates are frameworkreported when available; lack of standardized cache telemetry limits cross-runtime precision.

**Concurrency and batching.** Concurrency sweeps use synthetic open-loop clients and fixed sampling parameters. Real systems may include adaptive schedulers, user think-time, and backpressure. Apple-native runtimes currently lack standardized continuous batching, limiting generality to single-host micro-batch regimes.

**Reproducibility and artifacts.** We release scripts, logs, commit hashes, and plotting code to regenerate all figures. Differences in local toolchains (compilers, Homebrew, codesigning) can affect TVM/Metal builds (MLC-LLM) and wheel selection (MLX). We document exact commands and environment manifests to facilitate replication.

**External baselines.** Comparisons to NVIDIA A100+vLLM are contextual, not head-to-head on identical workloads in this paper. Reported A100 figures are used as an external ceiling for practicality; absolute ratios depend on model size, quantization, and scheduler settings.